\documentclass{aa}  

\usepackage{color}
\usepackage{subfigure}
\usepackage{graphicx}
\usepackage{epsfig}
\usepackage{hyperref}
\usepackage{longtable}
\newcommand{\plck}{PLCK\,G285.0-23.7~}
\newcommand{\plckp}{PLCK\,G285.0-23.7}

\usepackage{txfonts}
\begin{document}

   \title{ATCA observations of the MACS-Planck Radio Halo Cluster Project}
   \titlerunning{Radio halo in \plck}
   \subtitle{I. New detection of a radio halo in \plck}
   \authorrunning{Martinez Aviles et al.}
   \author{G. Martinez Aviles \inst{1} \fnmsep\thanks{Gerardo.Martinez-Aviles@oca.eu},
          C. Ferrari\inst{1},
          M. Johnston-Hollitt \inst{2}, 
          L. Pratley \inst{2}, 
          G. Macario\inst{1},
          T. Venturi \inst{3},
          G. Brunetti \inst{3},
          R. Cassano \inst{3},
          D. Dallacasa \inst{4,3}
          H.~T. Intema \inst{5},
          S. Giacintucci \inst{6},
          G. Hurier \inst{7},
          N. Aghanim \inst{7},
          M. Douspis \inst{7}
          \and
          M. Langer \inst{7}      
          }

   \institute{Laboratoire Lagrange, Universit\'e C\^ote d'Azur, Observatoire de la C\^ote d'Azur, CNRS, Blvd de l'Observatoire, CS 34229, 06304 Nice cedex 4, France
         \and
             School of Chemical \& Physical Sciences, Victoria University of Wellington, PO Box 600, Wellington 6140, New Zealand
                 \and
                 INAF- Osservatorio di Radioastronomia, via P. Gobetti 101, I-40129, Bologna, Italy
                 \and
                 Department of Physics and Astronomy, UniBO, via Ranzani 1, I-40127 Bologna, Italy
                 \and
                 Leiden Observatory, Leiden University, Niels Bohrweg 2, NL-2333CA, Leiden, The Netherlands
         \and
         Department of Astronomy, University of Maryland, College Park, MD 20742-2421, USA
         \and     
         Institut d'Astrophysique Spatiale, CNRS, Univ. Paris-Sud, Universit\'e Paris-Saclay, B\^at. 121, 91405 Orsay cedex, France
         }
   \date{Received April 25, 2016; accepted June 24, 2016}

 
  \abstract
   {}
   {We investigate the possible presence of diffuse radio emission in the intermediate redshift, massive cluster \plck ($z$=0.39, $M_{500}$ = 8.39 $\times$ $10^{14}M_{\odot}$).}
   {Our 16cm-band ATCA observations of \plck allow us to reach a rms noise level of $\sim$ 11 $\mu$Jy/beam on the wide-band (1.1--3.1 GHz), full-resolution ($\sim$ 5 arcsec) image of the cluster, making it one of the deepest ATCA images yet published. We also re-image visibilities at lower resolution in order to achieve a better sensitivity to low-surface-brightness extended radio sources.}
   {We detect one of the lowest luminosity radio halos known at z$>$0.35, characterised by a slight offset from the well-studied 1.4 GHz radio power vs. cluster mass correlation. Similarly to most known radio-loud clusters (i.e. those hosting diffuse non-thermal sources), \plck has a disturbed dynamical state. Our analysis reveals a similarly elongated X-ray and radio morphology. While the size of the radio halo in \plck is smaller than lower redshift radio-loud clusters in the same mass range, it shows a similar correlation with the cluster virial radius, as expected in the framework of hierarchical structure formation. } 
   {}


   \maketitle
%

\section{Introduction}

Radio halos (RHs) are Mpc-scale diffuse synchrotron sources observed in the central regions of galaxy clusters \citep[see e.g.][for a recent observational review]{2012A&ARv..20...54F}. Radio halos are found in about one-third of massive clusters \citep[see][for an updated study]{2015A&A...580A..97C} and are located in merging systems  (e.g. \citealt{2013ApJ...777..141C}; see \citealt{2014MNRAS.444L..44B} for a case of radio halo in an apparently relaxed system).

It is generally believed that turbulence induced by mergers in galaxy clusters can reaccelerate the relativistic electrons responsible for the origin of radio halos \citep[see e.g.][for reviews]{2001Brunetti,2001ApJ...557..560P,2014Brunetti}. This scenario naturally explains the connection between radio halos and mergers; however, it also poses fundamental questions on the micro-physics of the mechanisms that are responsible for the acceleration and transport of these relativistic electrons \citep[see][for recent discussions]{2015ApJ...800...60M,2016MNRAS.458.2584B}.

The study of the connection between radio halos and the thermal properties of the hosting clusters  sheds light on these mechanisms and in general on the interplay between thermal and non-thermal components in these systems.

Several correlations between thermal and  non-thermal properties of galaxy clusters have been found \citep[$P_{1.4}$--$L_{X}$, $P_{1.4}$--Mass, $P_{1.4}$--$Y_{500}$; where $P_{1.4}$ and $Y_{500}$ are,  respectively, the radio power of halos at 1.4 GHz and the cluster integrated SZ signal within the radius at which the mean mass density is 500 times the critical density at the cluster redshift $R_{500}$; see e.g.][]{2012MNRAS.421L.112B,2013ApJ...777..141C}. In this respect radio follow up of X-ray or mass-selected samples of galaxy clusters provide a unique way to probe the formation of radio halos and their connection with cluster mergers. Models predict that the bulk of radio halos should be generated at z=0.2-0.4 \citep[e.g.][]{2006Cassano}, yet current statistical studies are available only for a limited range of masses and redshifts \citep[e.g.][]{2015A&A...580A..97C}.  

In this framework, this work is part of the MACS-Planck
Radio Halo Cluster Project \citep[][]{2014A&A...565A..13M} conceived to explore the origin and occurrence of RHs and their connection with the dynamical state of the host systems by extending previous studies to a higher redshift range. \citet{2012A&ARv..20...54F} show that the redshift distribution of clusters hosting RHs is  homogeneous up to $z$=0.35, but statistically incomplete at higher redshifts. Our sample includes 32 intermediate redshift clusters (0.3 $<z<$ 0.45), which are being analysed through deep $\sim$ 325 MHz GMRT or $\sim$ 2.1 GHz ATCA observations, depending on the declination of the targets. In this paper we present the discovery of a RH in \plckp, which is a massive cluster \citep[$M_{500}$ = 8.39 $\times$ $10^{14}M_{\odot}$][]{2011A&A...536A...9P} located at redshift $z$=0.39. 

In the following, Sect.\,\ref{sect:data} describes the ATCA observations of \plck together with the data reduction and the image reconstruction strategy. The physical properties of the detected diffuse radio source and our analysis of its nature are presented in Sect.\,\ref{sect:results}. Our study is discussed and concluded in Sect. \,\ref{sect:DeC}. In the $\Lambda$CDM cosmology adopted throughout this paper (with ${\rm H}_{0}$=70 km s$^{-1}$ Mpc$^{-1}$, $\Omega_{M}$=0.3, $\Omega_{\Lambda}$=0.7), 1~arcsec corresponds to 5.29~kpc at the redshift of \plckp.


\section{Radio observations and data reduction}\label{sect:data}

   Radio observations of \plck were undertaken on the Australia Telescope Compact Array (ATCA) in three separate array configurations (6D, 750A, 1.5A) using the Compact Array Broadband Backend (CABB)  correlator with a central frequency of 2.1 GHz and spanning 1.1-3.1 GHz. Observations were carried out in continuum mode with the correlator set to produce 2000 $\times$ 1 MHz channels. Details of the observations can be found in Table \ref{tab:obs}. The primary flux scale was set relative to the unresolved source PKS\,B1934-638 for which the detailed spectral behaviour is well understood over the ATCA band. The phase calibration was initially performed relative to PKS B1036-697 for first of two observations in the 6D configuration which commenced on 8 June 2012. However, it was quickly noted that PKS\,B1036-697 was partially confused in this configuration and so this calibrator was replaced with PKS\,B0606-795 for the remaining observation run.

   \begin{table}
      \caption[]{Details of our ATCA observations towards \plck: dates of observations (Col. 1) with different array configurations (Col. 2); Observation time (Col. 3); phase calibrator (Col. 4). The observations were taken at a central frequency of 2.1 GHz and a bandwidth of 2 GHz}
         \label{tab:obs}
     $$ 
         \begin{tabular}{lllll}
            \hline
            \noalign{\smallskip}
            Date      &   Config.   & Obs. time   & Calibrator \\
                      &             &  (min.)     &            \\
            \noalign{\smallskip}
            \hline
            \noalign{\smallskip}
            2012-Jun-8   & 6D   & 704.0  & PKS\,B1036-697 \\
            2012-Jun-9   &  6D  & 523.3  & PKS\,B0606-795 \\
            2012-Jun-29   & 750A &  531.3 & PKS\,B0606-795 \\
            2013-Sep-6  & 1.5A & 803.6  & PKS\,B0606-795\\
            \noalign{\smallskip}
            \hline
         \end{tabular}
     $$ 
   \end{table}

\begin{figure}
   \centering
   \includegraphics[angle=-90, width=\hsize]{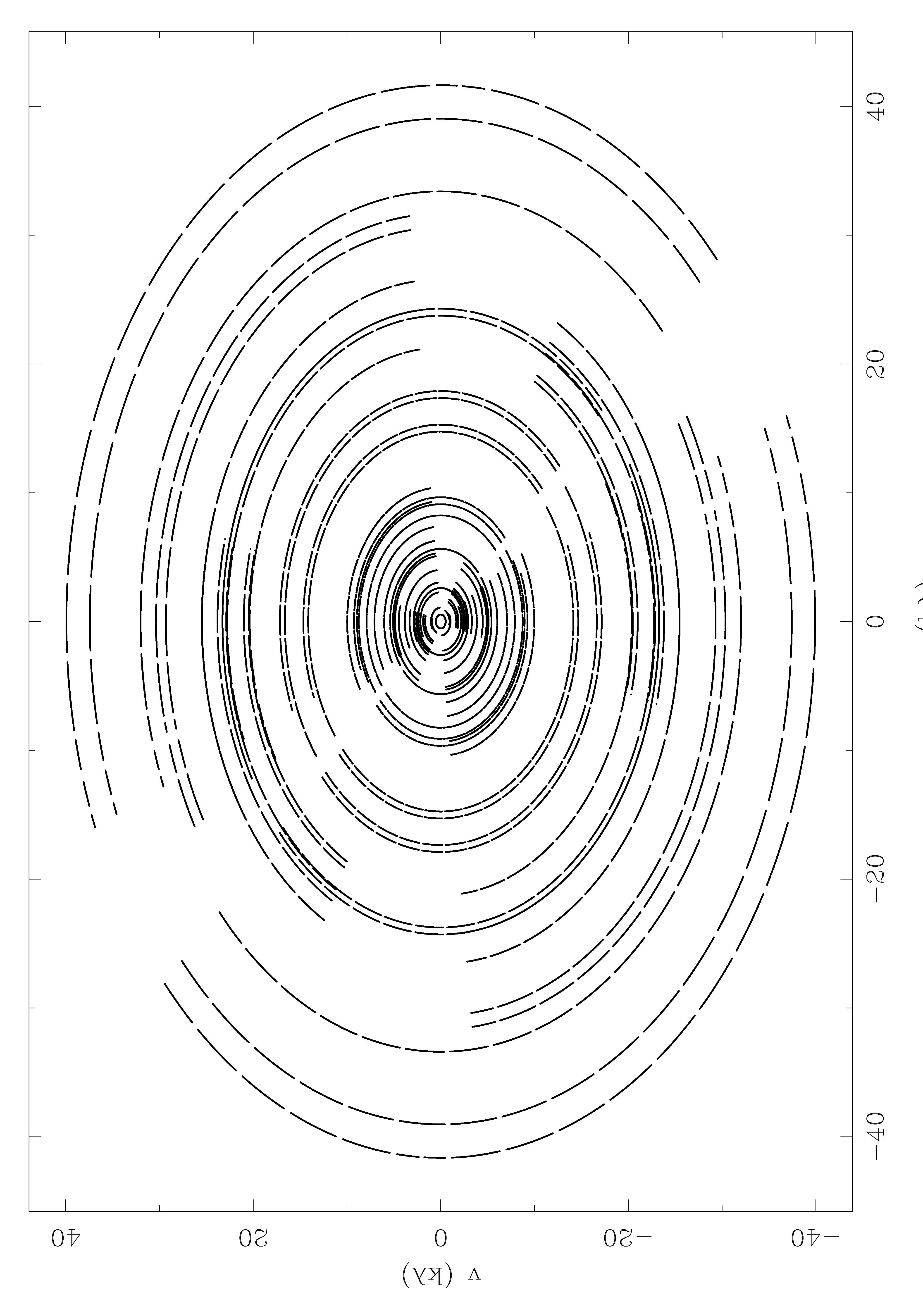}
      \caption{uv-coverage of \plck observations in the 500~MHz band centred at 1.867 MHz. The wide-band  of the full 2 GHz observations completes the coverage.}
         \label{Figuvcover}
   \end{figure}

   \begin{figure*}[ht]
   \centering
   \includegraphics[bb=60 233 540 622,width=15cm,clip]{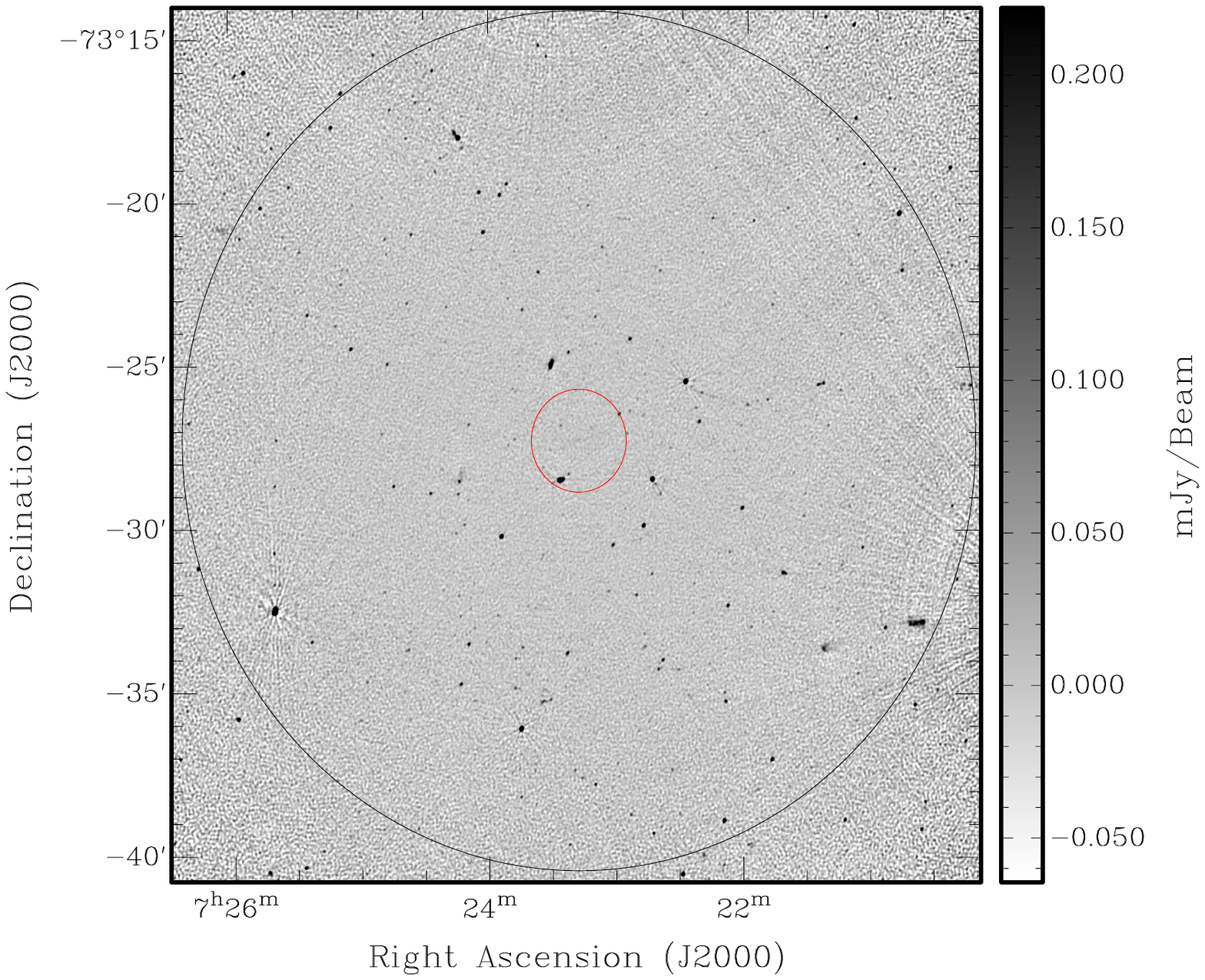}
      \caption{Final full-resolution, wide-band image of \plckp. The outer, big circle denotes the boundary of the primary beam with a radius of $\sim$0.22 degrees. The central 1 Mpc diameter region is indicated by the inner, smaller red circle. The rms noise of this image is 11.3 $\mu$Jy/beam  and the resolution is 5".
             }
         \label{fig:Figwideband}
   \end{figure*}

Radio frequency interference (RFI) and bad channels were manually excised from the target and primary and secondary calibrators using a combination of clipping algorithms and visual inspection via the PGFLAG task within MIRIAD \citep{1995ASPC...77..433S}.  Owing to the  nature of CABB data, it is necessary to perform calibration and cleaning on narrower frequency intervals. The precise intervals used depend on properties of the observations (phase stability) and the complexity of sources in the field, which affects the success of cleaning. We conducted a number of trials and determined that $\sim$500 MHz $\times$ 4 sub-bands produce the optimal results for these data. Thus, the target, primary, and secondary calibrator data were divided into the required sub-bands centred at 1.381 (from 1.130 GHz to 1.630 GHz), 1.867 (from 1631 GHz to 2.130 GHz), 2.380 (from 2.131 GHz to 2.630 GHz), and 2.769 MHz (from 2.631 GHz to 3.100 GHz). Each sub-band was then calibrated in MIRIAD as per standard procedures outlined in the 2014 MIRIAD manual to produce four continuum images. During this procedure it became apparent that data from the 6D array suffered from a bandpass ripple on all baselines associated with antennas 5 and 6 which persisted for 75 minutes. This effect is a transient error of unknown origin seen rarely in ATCA data and was not evident in data from  the 750A or 1.5A configuration. Attempts to correct for this proved unsuccessful and baselines to antennas 5 and 6 were removed for this configuration over this time. An example of the resultant uv-coverage for the combination of all three configurations for the 1.867 GHz sub-band image is shown in Fig.\,\ref{Figuvcover}.

 \begin{table}[]
    \caption[]{Properties of the full-resolution ({\it first five rows}) and tapered ({\it last five rows}) radio maps in different frequency bands centred at $\nu_0$. The band-width is $\sim$500 MHz for IDs from ``Block 1'' to ``Block 4''; it includes the total frequency range of observations (see Table\,\ref{tab:obs}) for the ``Wide-band'' ID.}
\centering
\label{table:maps}
$$\begin{tabular}{c  c  c  c  c }
\hline
Band ID       & $\nu_0$ &  RMS noise  & Beam Size  & PA  \\
              & (GHz) &  ($\mu$Jy/beam) & (''$\times$'') & (deg) \\
\hline 
\hline \\[-0.5em]
\multicolumn{5}{c}{Full-resolution images}\\[0.5em]
\hline
Block 1              &  2.769       & 20.1  & 2.46 $\times$ 2.09     &    -41.24    \\
Block 2              &  2.380       & 19.9  & 2.88 $\times$ 2.45     &    -41.52    \\
Block 3              &  1.867       & 18.1  & 3.62 $\times$ 3.10     &    -35.24    \\
Block 4              &  1.381       & 25.3  & 5.19 $\times$ 4.38     &    -42.31    \\
Wide-band            &  2.030       & 11.3  & 5.20 $\times$ 4.38     &    -42.31    \\    
\hline
\hline \\[-0.5em]
\multicolumn{5}{c}{Tapered images}\\[0.5em]
\hline
Block 1              &  2.769       & 31.6  & 30.54 $\times$ 19.69     &    83.92    \\
Block 2              &  2.380       & 38.1  & 30.46 $\times$ 20.75     &    -89.52    \\
Block 3              &  1.867       & 51.1  & 30.96 $\times$ 22.68     &    -81.17    \\
Block 4              &  1.381       & 71.3  & 31.03 $\times$ 23.93     &    -72.66    \\
Wide-band                    &  2.030       & 25.3 &  31.48 $\times$ 25.09      & -73.07 \\
\hline    
\hline
\end{tabular}$$
\end{table}
    
Following calibration, sub-band images were created out to the second null in the point spread function using the task INVERT and a Steer CLEAN \citep{1984A&A...137..159S} was applied to all sources within the primary beam. Two bright sources (at coordinates RA=07:19:25.285, Dec=-73:32:21.16 and RA=07:23:30.799, Dec=-73:08:40.16) with fluxes of just over 100 mJy sitting just outside of the primary beam were difficult to clean. They were modelled and subtracted from all of the visibility data and the sub-bands were re-imaged. Details of the resultant sub-band images are presented in Table \ref{table:maps}. The first 200 MHz of the band is strongly affected by RFI meaning the image at 1.3 GHz has the lowest sensitivity. 
   
The sub-band images were convolved with a Gaussian to a common resolution, slightly lower than the lowest resolution sub-band image. The sub-band images were then added together to create a final wide-band image  (see Table \ref{table:maps}). As the noise levels in the sub-band images are mostly identical for the three highest bands and only differed marginally in the lowest frequency band, it was not necessary to weight the images relative to the sensitivities in the mosaicing process. The final deep ATCA image achieves a root mean squared (rms) noise of $\sim$11 $\mu$Jy/beam measured with AIPS TVSTAT at the field centre in regions inside the primary beam without any trace of point sources or diffuse emission. We get a dynamic range of 10,000:1, making it one of the deepest, highest dynamic range ATCA images yet published, equal to the recent image of the Bullet Cluster at 11$\mu$Jy/beam \citep{2014MNRAS.440.2901S,2015MNRAS.449.1486S}. The image is shown in Fig.\,\ref{fig:Figwideband}.
   
\begin{figure*}
   \centering
   \includegraphics[width=0.49\hsize]{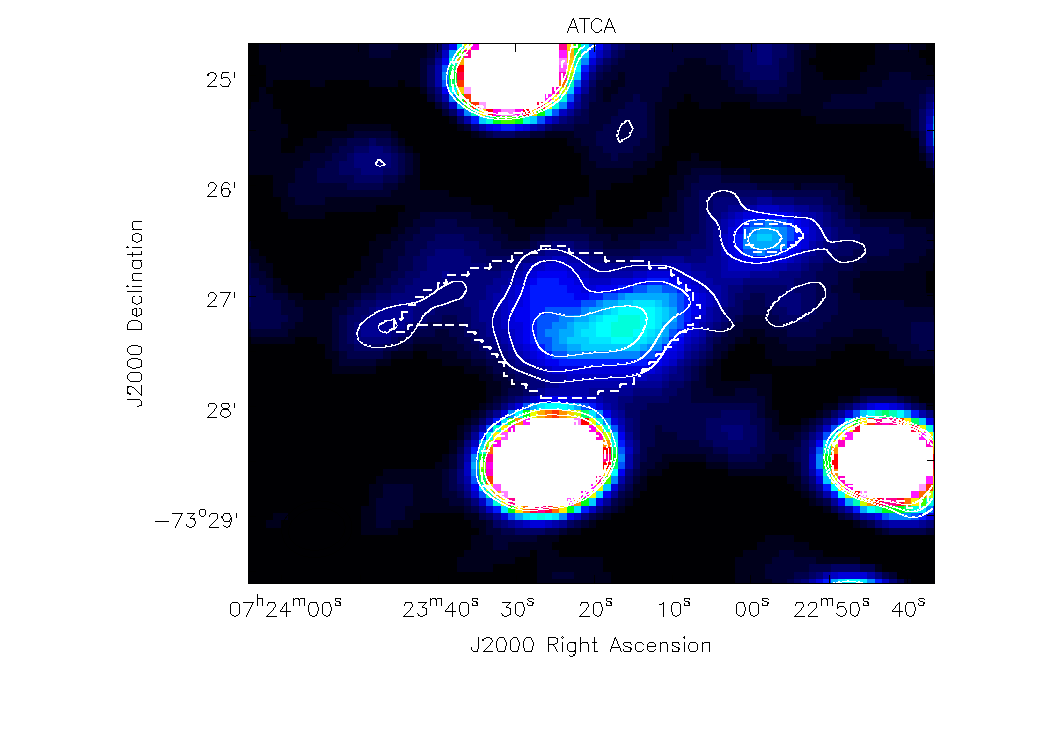}
   \includegraphics[width=0.49\hsize]{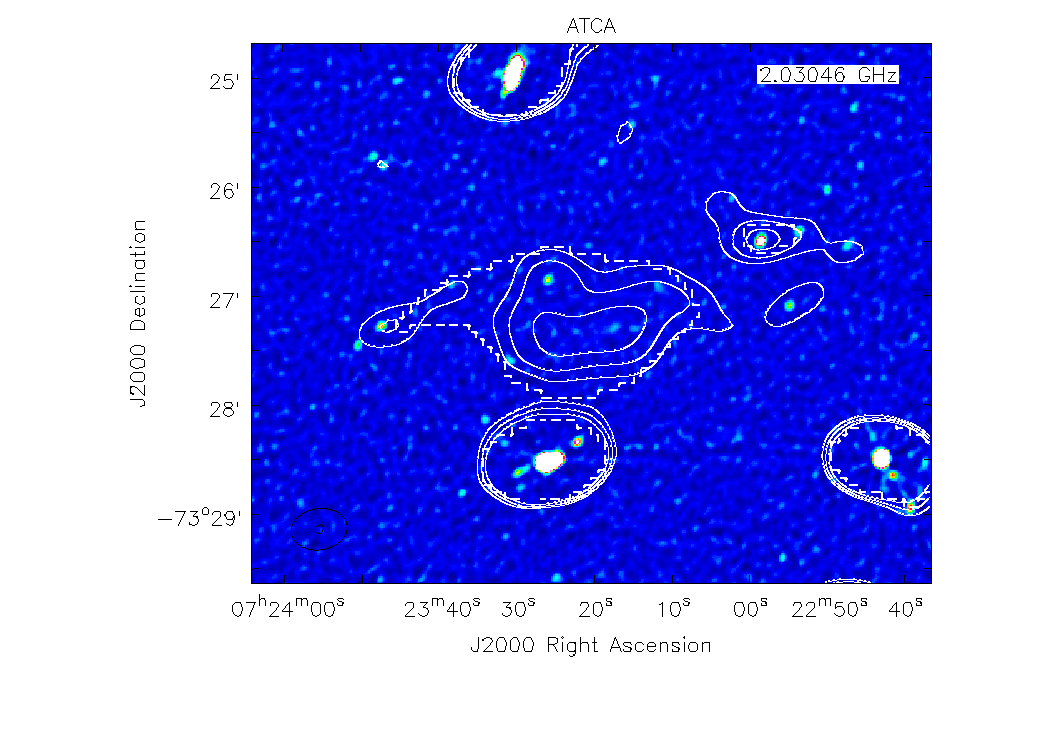}
      \caption{Contours of the ATCA tapered image centred at 1.867 GHz are overlaid on the ATCA wide-band tapered ({\it left}) and full-resolution  maps ({\it right}). Dashed contours correspond to the islands of significant emission detected by PyBDSM on the Block 3 tapered map (cyan contours in Fig.\ref{fig:pybdsm}). The continuous curves trace instead ``classical'' 3, 3 $\times \sqrt{2}$, 6 $\sigma$ contours (i.e. with the 51.1 $\mu$Jy/beam rms value of the map estimated by hand on the Block 3 tapered map at $\sim$ 30 arcsec resolution; see Table \ref{table:maps}).              }
         \label{fig:multiw}
   \end{figure*}

A visual inspection of our  full-resolution ATCA image of \plck shows no obvious presence of diffuse radio emission in the central area where the cluster lies (see Fig.\,\ref{fig:Figwideband} and Fig.\,\ref{fig:multiw} for a zoomed version on the cluster). In order to investigate the possible presence of low-surface-brightness diffuse sources in the cluster, visibilities were re-imaged with a robust=0.5 weighting and  a $\sim$ 9 k$\lambda$ Gaussian tapering was applied allowing a FWHM of $\approx$ 30 arcsec (see Table \ref{table:maps}). Again, data were imaged in each of the four sub-bands with the taper applied and then mosaicked into a single deep map (see Table\,\ref{table:maps}). A diffuse, low-surface-brightness source is detected coincident with the cluster core.  

\section{Results}\label{sect:results}

\subsection{Detection and characterisation of a radio halo in \plck}\label{sect:char}

\begin{figure}
   \centering
   \includegraphics[width=0.9\hsize]{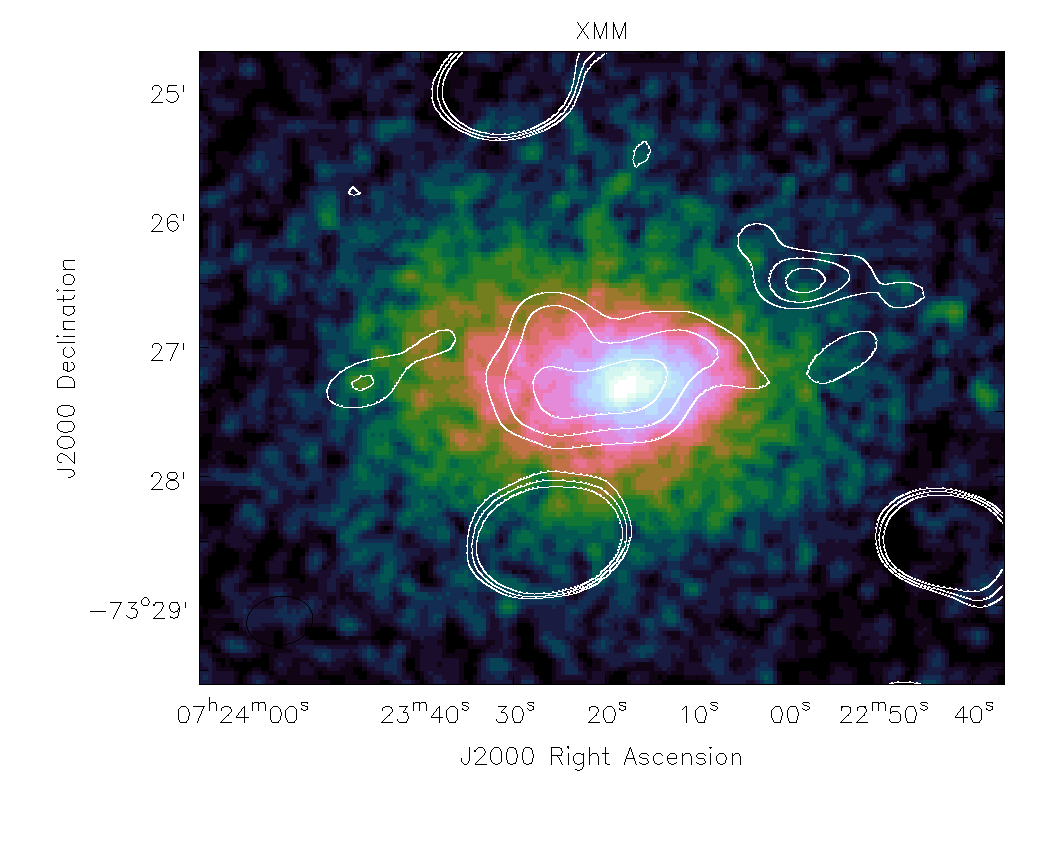}
      \caption{Radio contours of the ATCA tapered image centred at 1.867 GHz overlaid on the smoothed raw XMM image in the [0.3-2.0] keV energy band of \plck \citep{2011A&A...536A...9P}. Contour levels are the same as in Fig.\,\ref{fig:multiw}.}
         \label{fig:multib}
   \end{figure}

Contours of the low-resolution radio map of \plck are overlaid on the X-ray XMM image of the same sky region in Fig.\,\ref{fig:multib}. The morphology of the diffuse radio source matches very closely the thermal ICM X-ray emission in terms  of extension and of elongation (with a main axis along the east--west direction). This, together with the disturbed dynamical state of the cluster suggested by its X-ray morphology and density profile \citep{2011A&A...536A...9P}, allows us to confidently classify the diffuse source in \plck as a classical RH.   
   
\begin{table}[]
\centering

$$
\begin{tabular}{c  c  c  c}
\hline
Images ID & Angular size  & Physical size   & Flux \\
          & ('$\times$')  & (kpc $\times$ kpc) & (mJy)    \\
\hline
ID 1 &  2.34 $\times$ 1.34 &  742 $\times$ 425  &  2.02$\pm$0.25 \\
ID 2 &        "            &         "                &  2.11$\pm$0.26 \\
ID 2 ss &        "            &      "                   &  1.95$\pm$0.25 \\
\hline
\hline
ID 1 & 2.98 $\times$ 1.60 & 945 $\times$ 507  &  2.37$\pm$0.34  \\
ID 2 &              "     & "                       &  2.53$\pm$0.35 \\
ID 2 ss &   "                &  "                      &  2.17$\pm$0.44 \\
\hline
\end{tabular}
$$
\caption{Physical properties of the diffuse radio source at 1.867 GHz from tapered images. {\it Col. 1} gives the identification numbers of the tapered images on which we have performed the measurements: ID 1: Image with source subtraction in uv-data; ID 2: Image without source subtraction in uv-data; ID 2 ss: Image of ID2 after removing by hand the flux of point sources identified in the image plane. {\it Col. 2} and {\it Col. 3} correspond to the angular and physical size of the source, respectively (see text).  {\it Col. 4} reports the flux of the diffuse radio sources as measured with TVSTAT within the 3 $\sigma$ contours of the map with ID 2 ({\it top}), with TVSTAT within the 3 $\sigma$ contours calculated by PyBDSM for the map with ID 2 ({\it bottom}).}
\label{table:fluxes}
\end{table}
        
We now examine the radio flux density of the diffuse source on the tapered image relative to Block 3 (ID 2 in Table\,\ref{table:fluxes}), i.e. the image centred at 1.867 GHz and characterised by an average rms noise level of 51.1 $\mu$Jy/beam (Table \ref{table:maps}), as it shows the best sensitivity to the diffuse radio emission. Owing to the typical spectral behaviour of radio halos (with synchrotron spectral index $\alpha \gtrsim$1)\footnote{In this paper we use the convention S($\nu$) $\propto\nu^{-\alpha}$, with S($\nu$) being the radio flux density.}, the rms sensitivity of our maps is comparable from Block 1 to Block 3; however, a lower luminosity is expected for the diffuse radio source at the higher frequencies sampled by Blocks 1 and 2. In terms of sensitivity to steep synchrotron sources, Block 4 would have been the optimal frequency range in which to measure the radio halo flux density, if it had not been badly affected by RFI ($\sim$60\% of the data were flagged).

\begin{figure}
   \centering
   \includegraphics[bb=60 64 520 807,angle=-90,width=\hsize,clip]{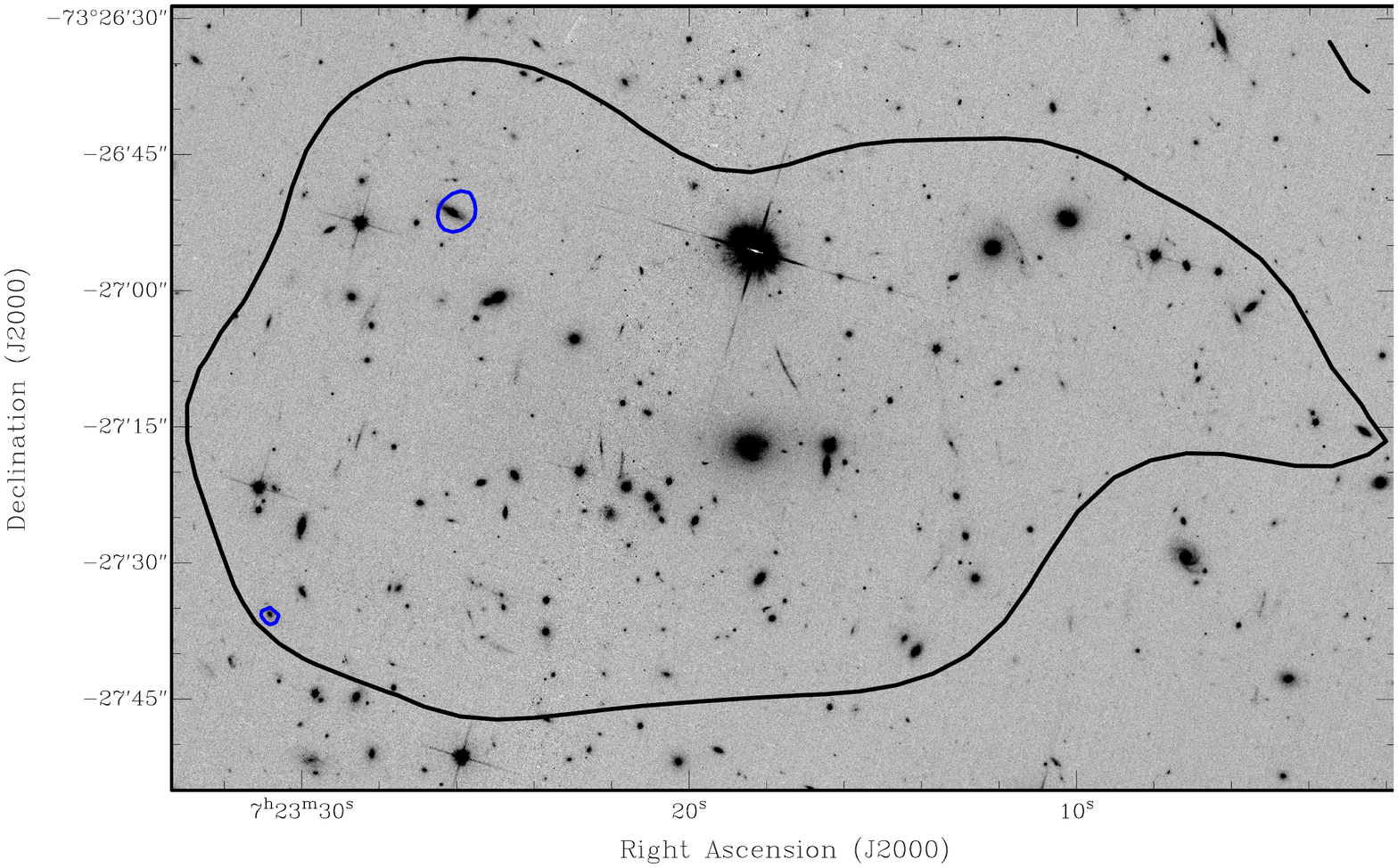}
      \caption{Zoom in the HST-ACS image of \plck central field (available in the HST archive). For reference, the 3 $\sigma$ contour of the ATCA tapered image centred at 1.867 GHz is shown in black. The blue contours indicate sources of significant radio emission in the full-resolution deep ATCA image of the cluster (i.e. 5 $\times$ rms of the final wide-band ATCA map shown in Fig.\,\ref{fig:Figwideband}).}
         \label{fig:hubble}
   \end{figure}

The total flux density was first measured on the Block 3 tapered image by integration over the source image within 3$\sigma$ contours using the AIPS task TVSTAT. By overlaying the 5$\sigma$ contours corresponding to the full-resolution wide-band radio map on the optical image of the cluster central region, we identify two point-like significant objects lying within the region occupied by the diffuse source on the low-resolution radio map. The two compact radio sources show quite clear optical counterparts (see Fig.\,\ref{fig:hubble}). We measured their flux densities using the AIPS verb JMFIT and we subtracted them from the total flux density measured inside the 3$\sigma$ contours of the tapered image.  

To verify that large-scale diffuse emission is not caused by the blending of discrete sources and to get a complementary flux measurement, we then produced an image of the diffuse cluster emission by subtracting compact sources in the uv-plane. We first identified the clean components of the discrete sources by using only the longest baselines (uv-range 3.6-40.0 k$\lambda$). These components were then subtracted from the original data set in the uv-plane by using the MIRIAD task UVMODEL. At this point, we calculated the flux density of the residual diffuse emission by integrating the surface brightness down to the same region considered before, i.e. within the $3\sigma$ level of the non-point source subtracted tapered map (ID 1 in Table \ref{table:fluxes}). The results of these flux density measurements are given in the top part of Table \ref{table:fluxes}. The error in flux density is calculated following the formula $\Delta F = \sqrt{(\sigma_{rms})^2N_{beam}+0.01F^{2}}$, where F is the measured flux density and $N_{beam}$ is the number of beams contained in the measured area. 

Table \ref{table:fluxes} shows the results of the different flux density measurements; all the values are consistent within the error bars. The flux measurement that we adopt in our following analysis is the one measured in the tapered map with point source subtracted from the uv-plane (ID1 in Table \ref{table:fluxes}), for which we obtain a flux density of 2.02$\pm$0.25 mJy at 1.867 GHz. 
 
In order to get an estimate of the size of the diffuse radio source, we measured the smallest ellipse that fully contains the 3$\sigma$ contours of the Block 3 tapered map without point source subtraction (ID2 in Table \ref{table:fluxes}). We obtain an angular extent of 2.34' $\times$ 1.34', which, for our cosmology, corresponds to a major and minor axis of $\sim$ 742 kpc and 425 kpc, respectively (see Table \ref{table:fluxes}). 

\begin{figure*}
   \centering
   \includegraphics[width=\hsize]{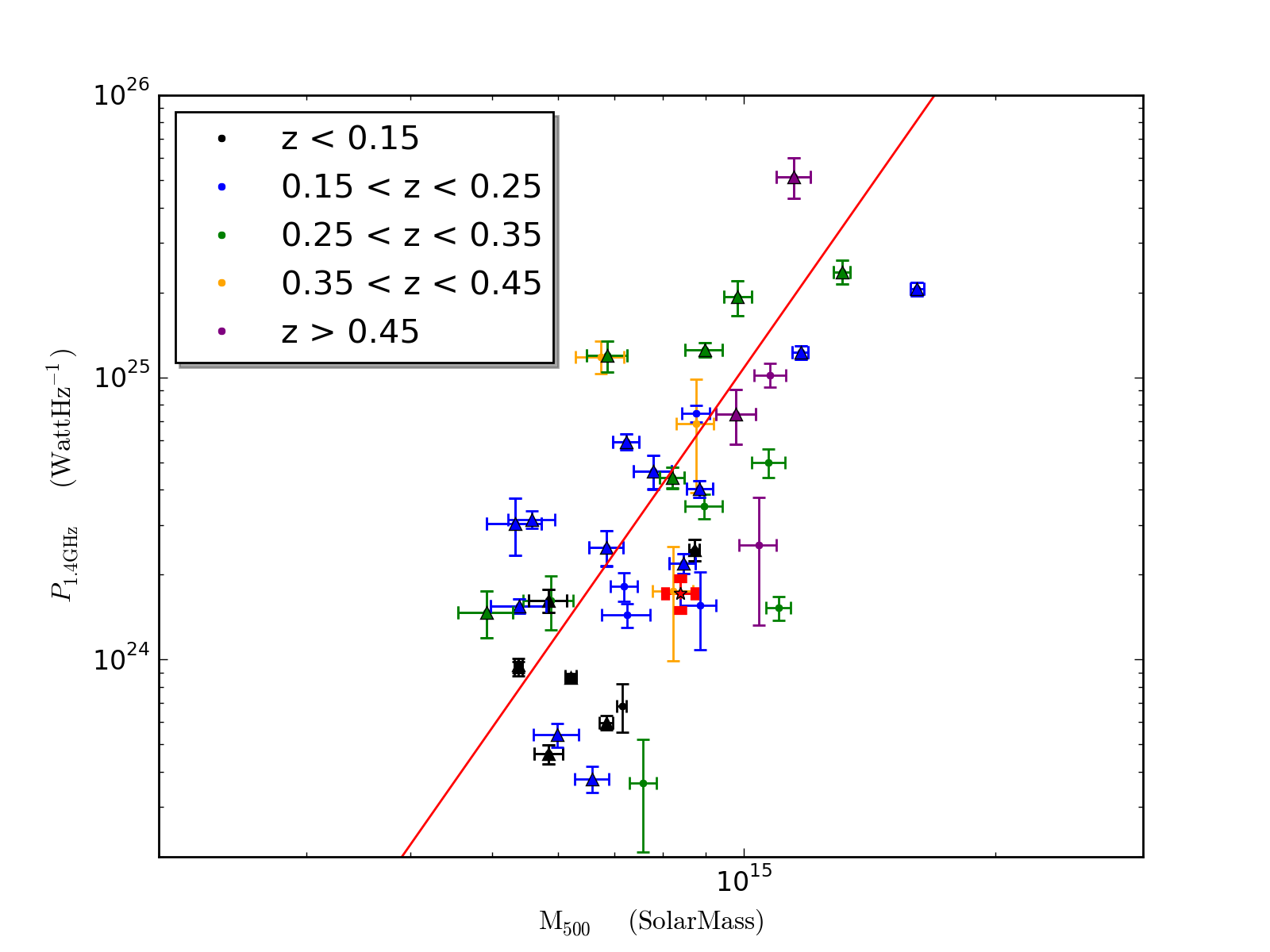}
      \caption{Synchrotron power of RHs at 1.4 GHz ($P_{1.4\,GHz}$) vs. cluster mass \citep[$M_{500}$ from][]{2015arXiv150201598P}. Symbols are colour-coded depending on the redshift of RH hosting clusters. The red line is the fit derived in the present paper, including only the RHs with radio fluxes measured at 1.4 GHz (indicated by triangles; see also Table \ref{tab:all_halo}). Analogously to \citet{2013ApJ...777..141C}, we exclude for the fitting ultra-steep spectrum RHs and halos detected in frequencies different from 1.4 GHz, both shown in dots (see Table \ref{tab:all_halo}). The red star corresponds to \plck detected in this work. When the spectral index is unknown (including \plck) we assume the value of $\alpha$=1.3.
             }
         \label{fig:Fighalos}
\end{figure*}

Finally, we measured with TVSTAT the total flux density on the ID1 and ID2 maps by following the contours of the significant islands of emission identified by the source finder software PyBDSM around the diffuse source (dashed contours in Fig.\,\ref{fig:multiw}; see Appendix\,\ref{subsect:PYBDSM} for details on the use of PyBDSM). The results obtained following this final method are given in the bottom part of Table\,\ref{table:fluxes}. 

\subsection{Radio luminosity vs. cluster mass correlation}\label{sect:radiomass}

We calculated the power of the newly discovered diffuse radio source and compared it with radio powers of previously discovered halos. We selected objects firmly classified as giant radio halos (i.e. with diffuse emission extending beyond the cluster core), whose flux density has been measured through radio interferometric observations and with a point source identification and subtraction strategy very similar to ours, and with information about $M_{500}$ in the PSZ2 cluster catalog \citep{2015arXiv150201598P}. The list of sources is presented in Table\,\ref{tab:all_halo}. 

In order to convert all values to the cosmology adopted in this paper and to take into account some inconsistencies found for published radio powers, we compiled the total flux densities of RHs reported in the literature (see Table \ref{tab:all_halo}). We then calculated radio powers using the formula $L_{\nu}=4\pi D_{L}^{2}S_{\nu}(1+z)^{\alpha-1}$, where $D_{L}$ is the luminosity distance of the cluster. Whenever a measured value of the spectral index is not available, we adopted $\alpha$=1.3. 

Traditionally RH powers are reported at 1.4 GHz; however, owing to the lower quality of the Block 4 image centred at 1.381 GHz, we considered the most reliable sub-band map (centred at 1.867 GHz) and then extrapolated to obtain the 1.4 GHz radio power. As the spectral index of the radio source is too uncertain over the ATCA band owing to the very low surface brightness of the object, we undertake this extrapolation assuming the value of $\alpha$=1.3. We obtain a value for the radio power at 1.4 GHz of 1.72 $\pm$ 0.22 $\times 10^{24}$ W/Hz. 

In Fig.\,\ref{fig:Fighalos} we plot the radio power {vs.} cluster mass for the RH in \plck (shown as a red star) as compared to all halos included in Table\,\ref{tab:all_halo}, which are indicated by triangles (colour-coded based on the redshift of their host clusters), except when they are classified as ultra steep spectrum (USS) radio halos and/or when their flux was not measured at $\approx$1.4 GHz (see Table\,\ref{tab:all_halo}), in which case they appear as dots (keeping the same colour code). Analogously to \citet{2013ApJ...777..141C}, we fit a power-law relation using linear regression in the log--log space and adopt a BCES--bisector regression algorithm \citep{1996ApJ...470..706A}. For the fitting, shown as a red line in Fig.\,\ref{fig:Fighalos} and characterised by a slope of $4.24\pm0.14$, we consider only the RHs represented as triangles in Fig.\,\ref{fig:Fighalos}. The best fit derived by \citet{2013ApJ...777..141C}, traditionally reported in the literature, has a slope of $3.77\pm0.57$. For completeness we also derived the value for the slope using all radio powers of halos present in Table \ref{tab:all_halo} for the best fit, which is $4.05\pm0.09$. The three results are consistent within the error bars. The RH in \plck is slightly under-luminous with respect to the best-fit relation, but is within the scatter of the observed data points. We also point out that, based on the present radio data, we do not find a clear trend with redshift of the $P_{1.4}-M_{500}$ correlation. 

\begin{figure}
   \centering
   \includegraphics[width=\hsize]{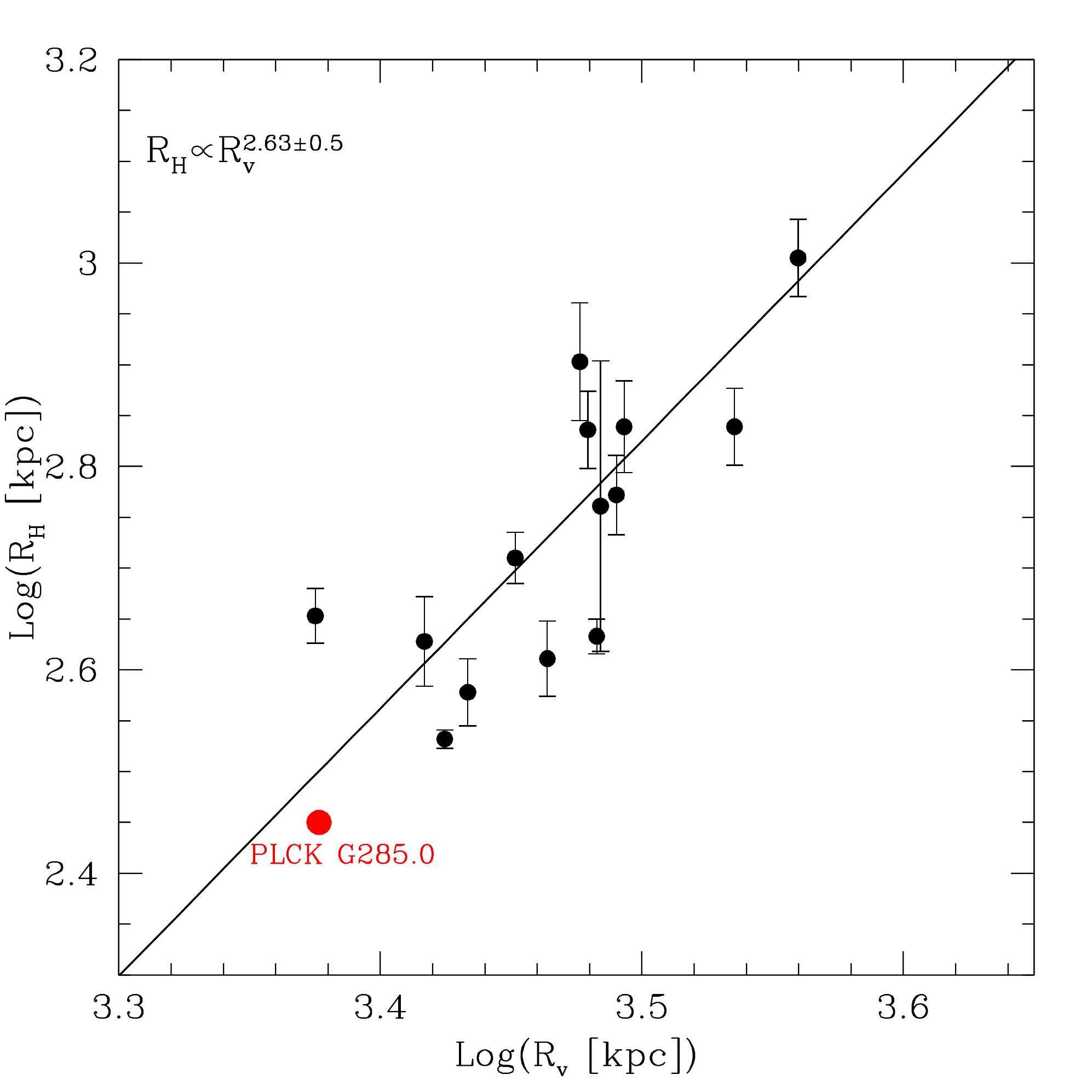}
      \caption{Correlation between radio halo sizes and the virial radius of their host clusters \citep[see][]{2007MNRAS.378.1565C}. The newly detected radio halo in \plck is indicated in red. 
             }
         \label{fig:sizeH}
\end{figure}

Following \citet{2007MNRAS.378.1565C}, we show in Fig.\,\ref{fig:sizeH} the position of the newly detected radio halo on the plot of radio halo sizes as a function of the virial radius of their host clusters. There is evidence that powerful radio halos are also bigger and hosted by more massive systems \citep[e.g.][]{2000NewA....5..335G,2001ApJ...548..639K,2007MNRAS.378.1565C,2009A&A...499..679M}. Although it is  a massive cluster ($M_{500} \sim 8.39 \times 10^{14}\,M_{\odot}$), \plck hosts a quite small radio halo of radius $\sqrt{a*b} \sim 281$ kpc, where a and b are the semi-major and semi-minor axis derived in Sect.\,\ref{sect:char}. However, if we consider the correlation between the size of the radio halos and the virial radius of their hosting clusters (Fig.\,\ref{fig:sizeH}), we see that \plck (red point) follows the trend of other radio loud clusters. This occurs because  \plck is at a relatively high redshift ($z\sim 0.39$), which means that its virial radius, and thus its radio halo, is smaller (owing to the cosmological growth and virialisation of galaxy clusters) than those of nearby clusters with the same mass.

\section{Discussion and conclusions}\label{sect:DeC}

We present high-sensitivity and wide-band (1.1 -- 3.1 GHz) ATCA observations of the galaxy cluster \plckp, which reveals the presence of a diffuse radio source with an extension (major axis) of $\approx$ 700 kpc and a total 1.4 GHz power of $\approx$1.7 $\times 10^{24}$ W/Hz, as extrapolated from measurements in the 500 MHz-wide sub-band centred at 1.867 GHz.

Both the diffuse nature of the source, which is not related to blending of compact radio objects, and the comparison of our high-sensitivity tapered radio map to the X-ray image of the cluster allow us to classify the detected emission as a classical RH. First, the morphology and size of the diffuse radio source indicate a very similar distribution of the non-thermal ICM and  the thermal component, traced by its X-ray bremsstrahlung radiation. Similarly to most known RHs, our newly detected halo is thus hosted by a massive cluster ($8.4 \times 10^{14} {\rm M}_{\odot}$), and there are  strong indications that it is a merging system. 

Our wide-band full-resolution radio map of \plck is one of the deepest ATCA images yet published, reaching an rms level of $\sim 11$ $\mu$Jy/beam with a  synthesised beam size of few arcsec, i.e. the kind of sensitivity expected for the upcoming continuum ASKAP-EMU survey \citep{2011PASA...28..215N}. Despite the depth and quality of our radio observations, the detection of the diffuse source required a tapered weighting (see Table\,\ref{table:maps}), which has allowed the detection of one of the lowest luminosity RH in the sub-sample of known radio-loud clusters at z$>$0.35 (Fig.\,\ref{fig:Fighalos}). This analysis is thus a very good example of the impact that the 30,000 square degrees (Dec<30$^{\circ}$) EMU survey will have on statistical studies of the non-thermal component of the ICM, if an appropriate data reduction strategy, which includes data tapering and compact source subtraction, is implemented.

The  measurements of the halo radio power can be significantly biased by several factors. First of all, the quality of our map centred at $\sim$1.4 GHz does not allow a proper detection and flux density measurement of the RH, which can instead be reliably measured in the sub-band image centred at $\sim$1.8 GHz. Since an in-band spectral index cannot be derived from our data, a value $\alpha$=1.3 is assumed here to get the k-corrected 1.4 GHz luminosity plotted in Fig.\,\ref{fig:Fighalos}. The other significant uncertainty in comparing our measurement to literature values is intrinsically related to the relative depth of the different published maps and to the method adopted for flux density measurements. Here we measure the flux density on images with point source subtraction both on the uv-data and on the image plane, showing that our different measurements are consistent within the error bars (see Table \ref{table:fluxes}). Is it worth mentioning that the telescope only measures power on the spatial scales of the sampled uv-coverage, which could limit the total power measured from the radio halo. This would be especially the case if the uv-coverage happened to lack at the physical scales of the radio halo.

With these elements in mind, the newly detected RH seems to be slightly under-luminous compared to objects hosted by clusters in a similar mass range, as shown in the ${\rm P}_{1.4~{\rm GHz}}$ vs. ${\rm M}_{500}$ plot of Fig.\,\ref{fig:Fighalos}; however,  it is not a clear outlier like the RH recently discovered by \citet{2015MNRAS.454.3391B}. 

It is expected that radio halos and their hosting clusters develop complex patterns with time in the radio--mass (or X-ray luminosity) diagram as a result of the evolution of cluster dynamics and particle acceleration and spectra \citep[e.g.][]{2013MNRAS.429.3564D}. For most of their lifetimes these systems are expected to be ``off-state'' (i.e. under-luminous) or in the region spanned by the correlation leading to an apparent bimodality \citep[e.g.][]{2009A&A...507..661B, 2013ApJ...777..141C}. In this respect under-luminous systems are expected to be generally associated with both young or old mergers \citep{2013MNRAS.429.3564D}; naively \plck could be in one of these stages. On the other hand the spectrum of radio halos also plays a role, with ultra-steep-spectrum radio halos, which are expected in less powerful mergers and at higher redshift (as a result of stronger IC losses), being statistically under-luminous \citep{2006Cassano, 2008Natur.455..944B, 2010Cassano} (see Fig.\,\ref{fig:Fighalos}). 

The new generation of more  sensitive radio telescopes will allow us to discriminate between these two possible scenarios in the case of \plck and similarly weak radio loud clusters through deeper multi-frequency radio observations.

\begin{acknowledgements}
We thank the referee for providing constructive comments and help in improving the contents of this paper. We warmly thank Gianluca Castignani for very useful discussions and Gabriel Pratt, Monique Arnaud, and Iacopo Bartalucci for supplying the X-ray images. GMA is supported by the Erasmus Mundus Joint Doctorate Program by Grants Number 2013-1471 from the agency EACEA of the European Commission. GMA, CF, and MJ-H acknowledge financial support from {\it Programme National Cosmologie et Galaxies (PNCG)} and {\it Universit\'e de Nice-Sophia Antipolis -- Programme Professeurs Invit\'es 2015}. MJ-H acknowledges the Marsden Fund administered by the Royal Society of New Zealand on behalf of the Ministry of Business, Innovation and Employment. LP is funded via Marsden Funding to MJ-H. The Australia Telescope Compact Array is part of the Australia Telescope National Facility, which is funded by the Australian Government for operation as a National Facility managed by CSIRO. The results of this paper are partially based on data retrieved from the ESA Planck Legacy Archive, NASA SkyView, and the Hubble Legacy Archive, which is a collaboration between the Space Telescope Science Institute (STScI/NASA), the Space Telescope European Coordinating Facility (ST-ECF/ESA), and the Canadian Astronomy Data Centre (CADC/NRC/CSA). 
\end{acknowledgements}

\begin{longtab}
\begin{longtable}{lcccccc}
\caption{\label{tab:all_halo} Collection of clusters known to host a giant radio halo (see Sect. \ref{sect:radiomass}). Clusters marked with $^{U}$ host USS radio halos. Col. 1: Cluster name; Col. 2: SZ mass proxy (M$_{500}$) from the last PSZ2 Planck cluster catalog \citep{2015arXiv150201598P}; Col. 3: Redshift from PSZ2 Planck cluster catalog \citep{2015arXiv150201598P}; Col. 4: Radio flux density at frequency given in Col. 5;  Col. 6: Radio Power at 1.4 GHz; Col. 7: References to the radio flux density.}\\
\hline\hline
 Name  & M$_{\rm SZ}$ &  z  & Flux density  & Freq.  & Power at 1.4 GHz & References  \\
       & ($\times 10^{14} {\rm M}_{\odot}$) &    & (mJy) & (GHz) & ($\times 10^{24} {W Hz}^{-1}$)& \\
\hline
\endfirsthead
\caption{continued.}\\
\hline\hline
 Name         & M$_{\rm SZ}$  &  z  & Flux density & Freq.  & Power at 1.4 GHz  & References  \\
              & ($\times 10^{14} {\rm M}_{\odot}$) & & (mJy) & (GHz) &($\times 10^{24} {W Hz}^{-1}$)& \\
\hline
\endhead
\hline
\endfoot
\\
\vspace{0.2cm}
A209            & 8.4642$^{+0.2837}_{-0.3160}$  & 0.206 &  16.9$\pm$1   & 1.4 &  2.19$\pm$0.17 & 1\\
\vspace{0.2cm}
A520            & 7.8003$^{+0.4033}_{-0.4131} $ & 0.203  & 34.4$\pm$1.5 & 1.4 &  4.65$\pm$0.63 & 2\\
\vspace{0.2cm}
A521$^{U}$      & 7.2556$^{+0.4743}_{-0.4887} $ & 0.2475  & 6.4$\pm$0.6 & 1.4 &  1.44$\pm$0.14 &3\\
\vspace{0.2cm}
A545            & 5.3940$^{+0.4059}_{-0.4098}$  & 0.154   &  23$\pm$1   & 1.4 & 1.55$\pm$0.09 & 4\\
\vspace{0.2cm}
A665            & 8.8590$^{+0.3230}_{-0.3202}$  & 0.1818  & 43.1$\pm$2.2 & 1.4 & 4.03$\pm$0.28 & 5 \\
\vspace{0.2cm}
A697$^{U}$      & 10.9984$^{+0.3716}_{-0.3671}$ & 0.282   & 5.2$\pm$0.5  & 1.4 & 1.53$\pm$0.15 & 6\\
\vspace{0.2cm}
A746            & 5.3352$^{+0.3917}_{-0.4028}$  & 0.2323   & 18$\pm$4 & 1.382  & 3.03$\pm$0.69 & 6 \\
\vspace{0.2cm}
A754$^{U}$      & 6.8539$^{+0.1249}_{-0.1277}$ & 0.0542    & 83$\pm$5 & 1.4    & 0.60$\pm$0.04 & 7\\
\vspace{0.2cm}
A773            & 6.8474$^{+0.3362}_{-0.3110}$ & 0.2172    & 12.7$\pm$1.3 & 1.4 & 2.50$\pm$0.36 & 2\\
\vspace{0.2cm}
A1300$^{U}$     & 8.9713$^{+0.4587}_{-0.4537}$ & 0.3075    & 130$\pm$10 & 0.325 & 3.55$\pm$0.35 & 8\\
\vspace{0.2cm}
A1351           & 6.8676$^{+0.3799}_{-0.3812}$ & 0.322     & 32.4$\pm$3.2 & 1.4 & 11.97$\pm$1.51 & 9\\
\vspace{0.2cm}
A1656           & 7.1652$^{+0.0674}_{-0.1073}$ & 0.0231    & 720$\pm$130 & 1.0 &  0.69$\pm$0.13 & 10\\
\vspace{0.2cm}
A1689           & 8.7689$^{+0.3368}_{-0.3368}$ & 0.1832    & 91.6$\pm$2.7 & 1.2 & 7.46$\pm$0.52 & 11\\
\vspace{0.2cm}
A1758           & 8.2173$^{+0.2727}_{-0.2824}$ & 0.2799    & 16.7$\pm$0.8 & 1.4 & 4.42$\pm$0.37 & 1\\
\vspace{0.2cm}
A1914           & 7.2358$^{+0.2582}_{-0.2612}$ & 0.1712    & 64$\pm$3     & 1.4 & 5.91$\pm$0.38 & 4\\
\vspace{0.2cm}
A1995           & 4.9242$^{+0.3773}_{-0.3691}$ & 0.3179    & 4.1$\pm$0.7  & 1.4 & 1.47$\pm$0.28 & 1 \\
\vspace{0.2cm}
A2034           & 5.8503$^{+0.2294}_{-0.2337}$ & 0.113     & 13.6$\pm$1.0 & 1.4 & 0.46$\pm$0.04  & 1\\
\vspace{0.2cm}
A2163         & 16.1164$^{+0.2968}_{-0.2922}$ & 0.203      & 155$\pm$2    & 1.4 & 20.58$\pm$1.17 & 12\\
\vspace{0.2cm}
A2219         & 11.6918$^{+0.2500}_{-0.2743}$ & 0.228      & 81$\pm$4     & 1.4  & 12.24$\pm$0.65 & 4 \\
\vspace{0.2cm}
A2254              & 5.5870$^{+0.3597}_{-0.3548}$ & 0.178  & 33.7$\pm$1.8 & 1.4  & 3.14$\pm$0.22 &  2 \\
\vspace{0.2cm}
A2255              & 5.3828$^{+0.0586}_{-0.0612}$ & 0.0809 & 56$\pm$3     & 1.4  & 0.93$\pm$0.06 & 13\\
\vspace{0.2cm}
A2256              & 6.2107$^{+0.1012}_{-0.0915}$ & 0.0581 & 103.4$\pm$1.1 & 1.4 & 0.87$\pm$0.02 & 14 \\
\vspace{0.2cm}
A2294              & 5.9829$^{+0.3671}_{-0.3741}$ & 0.178   & 5.8$\pm$0.5 & 1.4  & 0.54$\pm$0.05 & 1 \\
\vspace{0.2cm}
A2319              & 8.7351$^{+0.1132}_{-0.1240}$ & 0.0557  & 328$\pm$28  & 1.4  & 2.45$\pm$0.21 & 15\\
\vspace{0.2cm}
A2744              & 9.8356$^{+0.3947}_{-0.3754}$ & 0.3066  & 57.1$\pm$2.9 & 1.4 & 19.28$\pm$2.76 &2\\
\vspace{0.2cm}
A3411              & 6.5925$^{+0.3094}_{-0.3106}$ & 0.1687  &  4.8$\pm$0.5 & 1.4 & 0.38$\pm$0.04 &16 \\ 
\vspace{0.2cm}
A3888             &  7.1948$^{+0.2639}_{-0.2590}$  & 0.151 & 27.57$\pm$3.13 & 1.867 &1.89$\pm$0.22& 17 \\
\vspace{0.2cm}
PLCKG285.0-23.7 & 8.3925$^{+0.3332}_{-0.3404}$  & 0.39 & 2.02$\pm$0.25    & 1.867 & 1.72$\pm$0.22 & 18 \\
\vspace{0.2cm}
PLCKG171.9-40.7$^{U}$ & 10.7102$^{+0.4931}_{-0.4963}$ & 0.27 & 18$\pm$2 & 1.4 & 4.98$\pm$0.58 &19 \\
\vspace{0.2cm}
RXCJ0949.8+1708   & 8.2387$^{+0.4644}_{-0.4561}$  & 0.38 & 21.0$\pm$2.2 & 0.323  & 1.75$\pm$0.76 & 20\\
\vspace{0.2cm}
RXCJ0107.7+5408   & 5.8478$^{+0.3030}_{-0.3117}$  & 0.1066 & 55$\pm$5 & 1.382 & 1.62$\pm$0.15 & 6\\
\vspace{0.2cm}
RXCJ1514.9-1523 & 8.8607$^{+0.4054}_{-0.4578}$ & 0.2226 & 102$\pm$9 & 0.327   & 1.56$\pm$0.48 & 21\\
\vspace{0.2cm}
RXCJ2003.5-2323 & 8.9919$^{+0.4444}_{-0.4885}$ & 0.3171  & 35$\pm$2 & 1.4     & 12.54$\pm$0.75 & 9\\
\vspace{0.2cm}
MACSJ1149.5+2223$^{U}$ &10.4178$^{+0.5207}_{-0.5451}$ & 0.545 & 29$\pm$4 & 0.323    & 2.54$\pm$1.22 & 22\\
\vspace{0.2cm}
MACSJ0717.5+3745 &11.4871$^{+0.5347}_{-0.5482}$& 0.546& 41.5$\pm$4.1 & 1.4    & 51.45$\pm$8.43 & 23\\
\vspace{0.2cm}
MACSJ1752.0+4440& 6.7475$^{+0.4377}_{-0.4543}$ & 0.366 & 164$\pm$13  & 0.323  & 11.84$\pm$1.56 & 22 \\
\vspace{0.2cm}
MACSJ0553.4-3342   & 8.7720$^{+0.4399}_{-0.4629}$ & 0.431 & 62$\pm$5  & 0.323 & 6.87$\pm$2.96 & 22 \\   
\vspace{0.2cm}
CIZAJ1938.3+5409   & 7.5779$^{+0.2876}_{-0.2809}$   & 0.26  & 11.0$\pm$1.2  & 0.323 & 0.36$\pm$0.16 &20  \\
\vspace{0.2cm}
CL0016+16          & 9.7937$^{+0.5293}_{-0.5314}$   & 0.5456 & 5.5$\pm$0.5 & 1.4  & 7.43$\pm$1.63 & 5 \\
\vspace{0.2cm}
1E0657-56          & 13.1003$^{+0.2874}_{-0.2931}$ & 0.2965 & 78$\pm$5 & 1.4 & 23.68$\pm$2.30  & 24\\
\vspace{0.2cm}
El Gordo     & 10.7536$^{+0.4781}_{-0.4721}$ & 0.87 & 2.43$\pm$0.18    & 2.1 & 10.20$\pm$0.99 &25  \\
\end{longtable}
\tablebib{
(1) \citet{2009A&A...507.1257G}; (2) \citet{2001A&A...376..803G}; (3) \citet{2009ApJ...699.1288D}; (4) \citet{2003A&A...400..465B}; (5) \citet{2000NewA....5..335G}; (6) \citet{2011A&A...533A..35V}; (7) \citet{2011ApJ...728...82M}; (8) \citet{2013A&A...551A..24V}; (9) \citet{2009ApJ...704L..54G}; (10) \citet{1990ApJ...355...29K}; (11) \citet{2011A&A...535A..82V}; (12) \citet{2001A&A...373..106F}; (13) \citet{2005A&A...430L...5G}; (14) \citet{2006AN....327..553C}; (15) \citet{2013Farnsworth};(16) \citet{2013Vanweeren};(17) \citet{2016MNRAS.459.2525S}; (18) This paper; (19) \citet{2013Giacintucci}; (20) \citet{2015MNRAS.454.3391B}; (21) \citet{2011Giacintucci}; (22) \citet{2012Bonafede}; (23) \citet{2009A&A...505..991V}; (24) \citet{2000ApJ...544..686L}; (25) \citet{2014Lindner}.}
\end{longtab}

\bibliographystyle{aa}
\bibliography{plck285}

\begin{thebibliography}{53}
\expandafter\ifx\csname natexlab\endcsname\relax\def\natexlab#1{#1}\fi

\bibitem[{{Akritas} \& {Bershady}(1996)}]{1996ApJ...470..706A}
{Akritas}, M.~G. \& {Bershady}, M.~A. 1996, \apj, 470, 706

\bibitem[{{Bacchi} {et~al.}(2003){Bacchi}, {Feretti}, {Giovannini}, \&
  {Govoni}}]{2003A&A...400..465B}
{Bacchi}, M., {Feretti}, L., {Giovannini}, G., \& {Govoni}, F. 2003, \aap, 400,
  465

\bibitem[{{Basu}(2012)}]{2012MNRAS.421L.112B}
{Basu}, K. 2012, \mnras, 421, L112

\bibitem[{{Bonafede} {et~al.}(2012){Bonafede}, {Br{\"u}ggen}, {van Weeren},
  {Vazza}, {Giovannini}, {Ebeling}, {Edge}, {Hoeft}, \& {Klein}}]{2012Bonafede}
{Bonafede}, A., {Br{\"u}ggen}, M., {van Weeren}, R., {et~al.} 2012, \mnras,
  426, 40

\bibitem[{{Bonafede} {et~al.}(2015){Bonafede}, {Intema}, {Br{\"u}ggen},
  {Vazza}, {Basu}, {Sommer}, {Ebeling}, {de Gasperin}, {R{\"o}ttgering}, {van
  Weeren}, \& {Cassano}}]{2015MNRAS.454.3391B}
{Bonafede}, A., {Intema}, H., {Br{\"u}ggen}, M., {et~al.} 2015, \mnras, 454,
  3391

\bibitem[{{Bonafede} {et~al.}(2014){Bonafede}, {Intema}, {Br{\"u}ggen},
  {Russell}, {Ogrean}, {Basu}, {Sommer}, {van Weeren}, {Cassano}, {Fabian}, \&
  {R{\"o}ttgering}}]{2014MNRAS.444L..44B}
{Bonafede}, A., {Intema}, H.~T., {Br{\"u}ggen}, M., {et~al.} 2014, \mnras, 444,
  L44

\bibitem[{{Brunetti} {et~al.}(2009){Brunetti}, {Cassano}, {Dolag}, \&
  {Setti}}]{2009A&A...507..661B}
{Brunetti}, G., {Cassano}, R., {Dolag}, K., \& {Setti}, G. 2009, \aap, 507, 661

\bibitem[{{Brunetti} {et~al.}(2008){Brunetti}, {Giacintucci}, {Cassano},
  {Lane}, {Dallacasa}, {Venturi}, {Kassim}, {Setti}, {Cotton}, \&
  {Markevitch}}]{2008Natur.455..944B}
{Brunetti}, G., {Giacintucci}, S., {Cassano}, R., {et~al.} 2008, \nat, 455, 944

\bibitem[{{Brunetti} \& {Jones}(2014)}]{2014Brunetti}
{Brunetti}, G. \& {Jones}, T.~W. 2014, International Journal of Modern Physics
  D, 23, 30007

\bibitem[{{Brunetti} \& {Lazarian}(2016)}]{2016MNRAS.458.2584B}
{Brunetti}, G. \& {Lazarian}, A. 2016, \mnras, 458, 2584

\bibitem[{{Brunetti} {et~al.}(2001){Brunetti}, {Setti}, {Feretti}, \&
  {Giovannini}}]{2001Brunetti}
{Brunetti}, G., {Setti}, G., {Feretti}, L., \& {Giovannini}, G. 2001, \mnras,
  320, 365

\bibitem[{{Cassano} {et~al.}(2006){Cassano}, {Brunetti}, \&
  {Setti}}]{2006Cassano}
{Cassano}, R., {Brunetti}, G., \& {Setti}, G. 2006, \mnras, 369, 1577

\bibitem[{{Cassano} {et~al.}(2007){Cassano}, {Brunetti}, {Setti}, {Govoni}, \&
  {Dolag}}]{2007MNRAS.378.1565C}
{Cassano}, R., {Brunetti}, G., {Setti}, G., {Govoni}, F., \& {Dolag}, K. 2007,
  \mnras, 378, 1565

\bibitem[{{Cassano} {et~al.}(2013){Cassano}, {Ettori}, {Brunetti},
  {Giacintucci}, {Pratt}, {Venturi}, {Kale}, {Dolag}, \&
  {Markevitch}}]{2013ApJ...777..141C}
{Cassano}, R., {Ettori}, S., {Brunetti}, G., {et~al.} 2013, \apj, 777, 141

\bibitem[{{Cassano} {et~al.}(2010){Cassano}, {Ettori}, {Giacintucci},
  {Brunetti}, {Markevitch}, {Venturi}, \& {Gitti}}]{2010Cassano}
{Cassano}, R., {Ettori}, S., {Giacintucci}, S., {et~al.} 2010, \apjl, 721, L82

\bibitem[{{Clarke} \& {Ensslin}(2006)}]{2006AN....327..553C}
{Clarke}, T.~E. \& {Ensslin}, T. 2006, Astronomische Nachrichten, 327, 553

\bibitem[{{Cuciti} {et~al.}(2015){Cuciti}, {Cassano}, {Brunetti}, {Dallacasa},
  {Kale}, {Ettori}, \& {Venturi}}]{2015A&A...580A..97C}
{Cuciti}, V., {Cassano}, R., {Brunetti}, G., {et~al.} 2015, \aap, 580, A97

\bibitem[{{Dallacasa} {et~al.}(2009){Dallacasa}, {Brunetti}, {Giacintucci},
  {Cassano}, {Venturi}, {Macario}, {Kassim}, {Lane}, \&
  {Setti}}]{2009ApJ...699.1288D}
{Dallacasa}, D., {Brunetti}, G., {Giacintucci}, S., {et~al.} 2009, \apj, 699,
  1288

\bibitem[{{Donnert} {et~al.}(2013){Donnert}, {Dolag}, {Brunetti}, \&
  {Cassano}}]{2013MNRAS.429.3564D}
{Donnert}, J., {Dolag}, K., {Brunetti}, G., \& {Cassano}, R. 2013, \mnras, 429,
  3564

\bibitem[{{Farnsworth} {et~al.}(2013){Farnsworth}, {Rudnick}, {Brown}, \&
  {Brunetti}}]{2013Farnsworth}
{Farnsworth}, D., {Rudnick}, L., {Brown}, S., \& {Brunetti}, G. 2013, \apj,
  779, 189

\bibitem[{{Feretti} {et~al.}(2001){Feretti}, {Fusco-Femiano}, {Giovannini}, \&
  {Govoni}}]{2001A&A...373..106F}
{Feretti}, L., {Fusco-Femiano}, R., {Giovannini}, G., \& {Govoni}, F. 2001,
  \aap, 373, 106

\bibitem[{{Feretti} {et~al.}(2012){Feretti}, {Giovannini}, {Govoni}, \&
  {Murgia}}]{2012A&ARv..20...54F}
{Feretti}, L., {Giovannini}, G., {Govoni}, F., \& {Murgia}, M. 2012, \aapr, 20,
  54

\bibitem[{{Giacintucci} {et~al.}(2011){Giacintucci}, {Dallacasa}, {Venturi},
  {Brunetti}, {Cassano}, {Markevitch}, \& {Athreya}}]{2011Giacintucci}
{Giacintucci}, S., {Dallacasa}, D., {Venturi}, T., {et~al.} 2011, \aap, 534,
  A57

\bibitem[{{Giacintucci} {et~al.}(2013){Giacintucci}, {Kale}, {Wik}, {Venturi},
  \& {Markevitch}}]{2013Giacintucci}
{Giacintucci}, S., {Kale}, R., {Wik}, D.~R., {Venturi}, T., \& {Markevitch}, M.
  2013, \apj, 766, 18

\bibitem[{{Giacintucci} {et~al.}(2009){Giacintucci}, {Venturi}, {Cassano},
  {Dallacasa}, \& {Brunetti}}]{2009ApJ...704L..54G}
{Giacintucci}, S., {Venturi}, T., {Cassano}, R., {Dallacasa}, D., \&
  {Brunetti}, G. 2009, \apjl, 704, L54

\bibitem[{{Giovannini} {et~al.}(2009){Giovannini}, {Bonafede}, {Feretti},
  {Govoni}, {Murgia}, {Ferrari}, \& {Monti}}]{2009A&A...507.1257G}
{Giovannini}, G., {Bonafede}, A., {Feretti}, L., {et~al.} 2009, \aap, 507, 1257

\bibitem[{{Giovannini} \& {Feretti}(2000)}]{2000NewA....5..335G}
{Giovannini}, G. \& {Feretti}, L. 2000, \na, 5, 335

\bibitem[{{Govoni} {et~al.}(2001){Govoni}, {Feretti}, {Giovannini},
  {B{\"o}hringer}, {Reiprich}, \& {Murgia}}]{2001A&A...376..803G}
{Govoni}, F., {Feretti}, L., {Giovannini}, G., {et~al.} 2001, \aap, 376, 803

\bibitem[{{Govoni} {et~al.}(2005){Govoni}, {Murgia}, {Feretti}, {Giovannini},
  {Dallacasa}, \& {Taylor}}]{2005A&A...430L...5G}
{Govoni}, F., {Murgia}, M., {Feretti}, L., {et~al.} 2005, \aap, 430, L5

\bibitem[{{Kempner} \& {Sarazin}(2001)}]{2001ApJ...548..639K}
{Kempner}, J.~C. \& {Sarazin}, C.~L. 2001, \apj, 548, 639

\bibitem[{{Kim} {et~al.}(1990){Kim}, {Kronberg}, {Dewdney}, \&
  {Landecker}}]{1990ApJ...355...29K}
{Kim}, K.-T., {Kronberg}, P.~P., {Dewdney}, P.~E., \& {Landecker}, T.~L. 1990,
  \apj, 355, 29

\bibitem[{{Liang} {et~al.}(2000){Liang}, {Hunstead}, {Birkinshaw}, \&
  {Andreani}}]{2000ApJ...544..686L}
{Liang}, H., {Hunstead}, R.~W., {Birkinshaw}, M., \& {Andreani}, P. 2000, \apj,
  544, 686

\bibitem[{{Lindner} {et~al.}(2014){Lindner}, {Baker}, {Hughes}, {Battaglia},
  {Gupta}, {Knowles}, {Marriage}, {Menanteau}, {Moodley}, {Reese}, \&
  {Srianand}}]{2014Lindner}
{Lindner}, R.~R., {Baker}, A.~J., {Hughes}, J.~P., {et~al.} 2014, \apj, 786, 49

\bibitem[{{Macario} {et~al.}(2014){Macario}, {Intema}, {Ferrari}, {Bourdin},
  {Giacintucci}, {Venturi}, {Mazzotta}, {Bartalucci}, {Johnston-Hollitt},
  {Cassano}, {Dallacasa}, {Pratt}, {Kale}, \& {Brown}}]{2014A&A...565A..13M}
{Macario}, G., {Intema}, H.~T., {Ferrari}, C., {et~al.} 2014, \aap, 565, A13

\bibitem[{{Macario} {et~al.}(2011){Macario}, {Markevitch}, {Giacintucci},
  {Brunetti}, {Venturi}, \& {Murray}}]{2011ApJ...728...82M}
{Macario}, G., {Markevitch}, M., {Giacintucci}, S., {et~al.} 2011, \apj, 728,
  82

\bibitem[{{Miniati}(2015)}]{2015ApJ...800...60M}
{Miniati}, F. 2015, \apj, 800, 60

\bibitem[{{Mohan} \& {Rafferty}(2015)}]{2015ascl.soft02007M}
{Mohan}, N. \& {Rafferty}, D. 2015, {PyBDSM: Python Blob Detection and Source
  Measurement}, Astrophysics Source Code Library

\bibitem[{{Murgia} {et~al.}(2009){Murgia}, {Govoni}, {Markevitch}, {Feretti},
  {Giovannini}, {Taylor}, \& {Carretti}}]{2009A&A...499..679M}
{Murgia}, M., {Govoni}, F., {Markevitch}, M., {et~al.} 2009, \aap, 499, 679

\bibitem[{{Norris} {et~al.}(2013){Norris}, {Afonso}, {Bacon}, {Beck}, {Bell},
  {Beswick}, {Best}, {Bhatnagar}, {Bonafede}, {Brunetti}, {Budav{\'a}ri},
  {Cassano}, {Condon}, {Cress}, {Dabbech}, {Feain}, {Fender}, {Ferrari},
  {Gaensler}, {Giovannini}, {Haverkorn}, {Heald}, {Van der Heyden}, {Hopkins},
  {Jarvis}, {Johnston-Hollitt}, {Kothes}, {Van Langevelde}, {Lazio}, {Mao},
  {Mart{\'{\i}}nez-Sansigre}, {Mary}, {Mcalpine}, {Middelberg}, {Murphy},
  {Padovani}, {Paragi}, {Prandoni}, {Raccanelli}, {Rigby}, {Roseboom},
  {R{\"o}ttgering}, {Sabater}, {Salvato}, {Scaife}, {Schilizzi}, {Seymour},
  {Smith}, {Umana}, {Zhao}, \& {Zinn}}]{2013PASA...30...20N}
{Norris}, R.~P., {Afonso}, J., {Bacon}, D., {et~al.} 2013, \pasa, 30, e020

\bibitem[{{Norris} {et~al.}(2011){Norris}, {Hopkins}, {Afonso}, {Brown},
  {Condon}, {Dunne}, {Feain}, {Hollow}, {Jarvis}, {Johnston-Hollitt}, {Lenc},
  {Middelberg}, {Padovani}, {Prandoni}, {Rudnick}, {Seymour}, {Umana},
  {Andernach}, {Alexander}, {Appleton}, {Bacon}, {Banfield}, {Becker}, {Brown},
  {Ciliegi}, {Jackson}, {Eales}, {Edge}, {Gaensler}, {Giovannini}, {Hales},
  {Hancock}, {Huynh}, {Ibar}, {Ivison}, {Kennicutt}, {Kimball}, {Koekemoer},
  {Koribalski}, {L{\'o}pez-S{\'a}nchez}, {Mao}, {Murphy}, {Messias},
  {Pimbblet}, {Raccanelli}, {Randall}, {Reiprich}, {Roseboom},
  {R{\"o}ttgering}, {Saikia}, {Sharp}, {Slee}, {Smail}, {Thompson}, {Urquhart},
  {Wall}, \& {Zhao}}]{2011PASA...28..215N}
{Norris}, R.~P., {Hopkins}, A.~M., {Afonso}, J., {et~al.} 2011, \pasa, 28, 215

\bibitem[{{Petrosian}(2001)}]{2001ApJ...557..560P}
{Petrosian}, V. 2001, \apj, 557, 560

\bibitem[{{Planck Collaboration} {et~al.}(2015){Planck Collaboration}, {Ade},
  {Aghanim}, {Arnaud}, {Ashdown}, {Aumont}, {Baccigalupi}, {Banday},
  {Barreiro}, {Barrena}, \& et~al.}]{2015arXiv150201598P}
{Planck Collaboration}, {Ade}, P.~A.~R., {Aghanim}, N., {et~al.} 2015, ArXiv
  e-prints [\eprint[arXiv]{1502.01598}]

\bibitem[{{Planck Collaboration} {et~al.}(2011){Planck Collaboration},
  {Aghanim}, {Arnaud}, {Ashdown}, {Aumont}, {Baccigalupi}, {Balbi}, {Banday},
  {Barreiro}, {Bartelmann}, \& et~al.}]{2011A&A...536A...9P}
{Planck Collaboration}, {Aghanim}, N., {Arnaud}, M., {et~al.} 2011, \aap, 536,
  A9

\bibitem[{{Sault} {et~al.}(1995){Sault}, {Teuben}, \&
  {Wright}}]{1995ASPC...77..433S}
{Sault}, R.~J., {Teuben}, P.~J., \& {Wright}, M.~C.~H. 1995, in Astronomical
  Society of the Pacific Conference Series, Vol.~77, Astronomical Data Analysis
  Software and Systems IV, ed. R.~A. {Shaw}, H.~E. {Payne}, \& J.~J.~E.
  {Hayes}, 433

\bibitem[{{Shakouri} {et~al.}(2016){Shakouri}, {Johnston-Hollitt}, \&
  {Pratt}}]{2016MNRAS.459.2525S}
{Shakouri}, S., {Johnston-Hollitt}, M., \& {Pratt}, G.~W. 2016, \mnras, 459,
  2525

\bibitem[{{Shimwell} {et~al.}(2014){Shimwell}, {Brown}, {Feain}, {Feretti},
  {Gaensler}, \& {Lage}}]{2014MNRAS.440.2901S}
{Shimwell}, T.~W., {Brown}, S., {Feain}, I.~J., {et~al.} 2014, \mnras, 440,
  2901

\bibitem[{{Shimwell} {et~al.}(2015){Shimwell}, {Markevitch}, {Brown},
  {Feretti}, {Gaensler}, {Johnston-Hollitt}, {Lage}, \&
  {Srinivasan}}]{2015MNRAS.449.1486S}
{Shimwell}, T.~W., {Markevitch}, M., {Brown}, S., {et~al.} 2015, \mnras, 449,
  1486

\bibitem[{{Steer} {et~al.}(1984){Steer}, {Dewdney}, \&
  {Ito}}]{1984A&A...137..159S}
{Steer}, D.~G., {Dewdney}, P.~E., \& {Ito}, M.~R. 1984, \aap, 137, 159

\bibitem[{{Vacca} {et~al.}(2011){Vacca}, {Govoni}, {Murgia}, {Giovannini},
  {Feretti}, {Tugnoli}, {Verheijen}, \& {Taylor}}]{2011A&A...535A..82V}
{Vacca}, V., {Govoni}, F., {Murgia}, M., {et~al.} 2011, \aap, 535, A82

\bibitem[{{van Weeren} {et~al.}(2011){van Weeren}, {Br{\"u}ggen},
  {R{\"o}ttgering}, {Hoeft}, {Nuza}, \& {Intema}}]{2011A&A...533A..35V}
{van Weeren}, R.~J., {Br{\"u}ggen}, M., {R{\"o}ttgering}, H.~J.~A., {et~al.}
  2011, \aap, 533, A35

\bibitem[{{van Weeren} {et~al.}(2013){van Weeren}, {Fogarty}, {Jones},
  {Forman}, {Clarke}, {Br{\"u}ggen}, {Kraft}, {Lal}, {Murray}, \&
  {R{\"o}ttgering}}]{2013Vanweeren}
{van Weeren}, R.~J., {Fogarty}, K., {Jones}, C., {et~al.} 2013, \apj, 769, 101

\bibitem[{{van Weeren} {et~al.}(2009){van Weeren}, {R{\"o}ttgering},
  {Br{\"u}ggen}, \& {Cohen}}]{2009A&A...505..991V}
{van Weeren}, R.~J., {R{\"o}ttgering}, H.~J.~A., {Br{\"u}ggen}, M., \& {Cohen},
  A. 2009, \aap, 505, 991

\bibitem[{{Venturi} {et~al.}(2013){Venturi}, {Giacintucci}, {Dallacasa},
  {Cassano}, {Brunetti}, {Macario}, \& {Athreya}}]{2013A&A...551A..24V}
{Venturi}, T., {Giacintucci}, S., {Dallacasa}, D., {et~al.} 2013, \aap, 551,
  A24

\end{thebibliography}

\begin{appendix} 
\section{Detection of the source using PyBDSM}\label{subsect:PYBDSM}

\begin{figure*}
   \centering
   \includegraphics[width=\hsize]{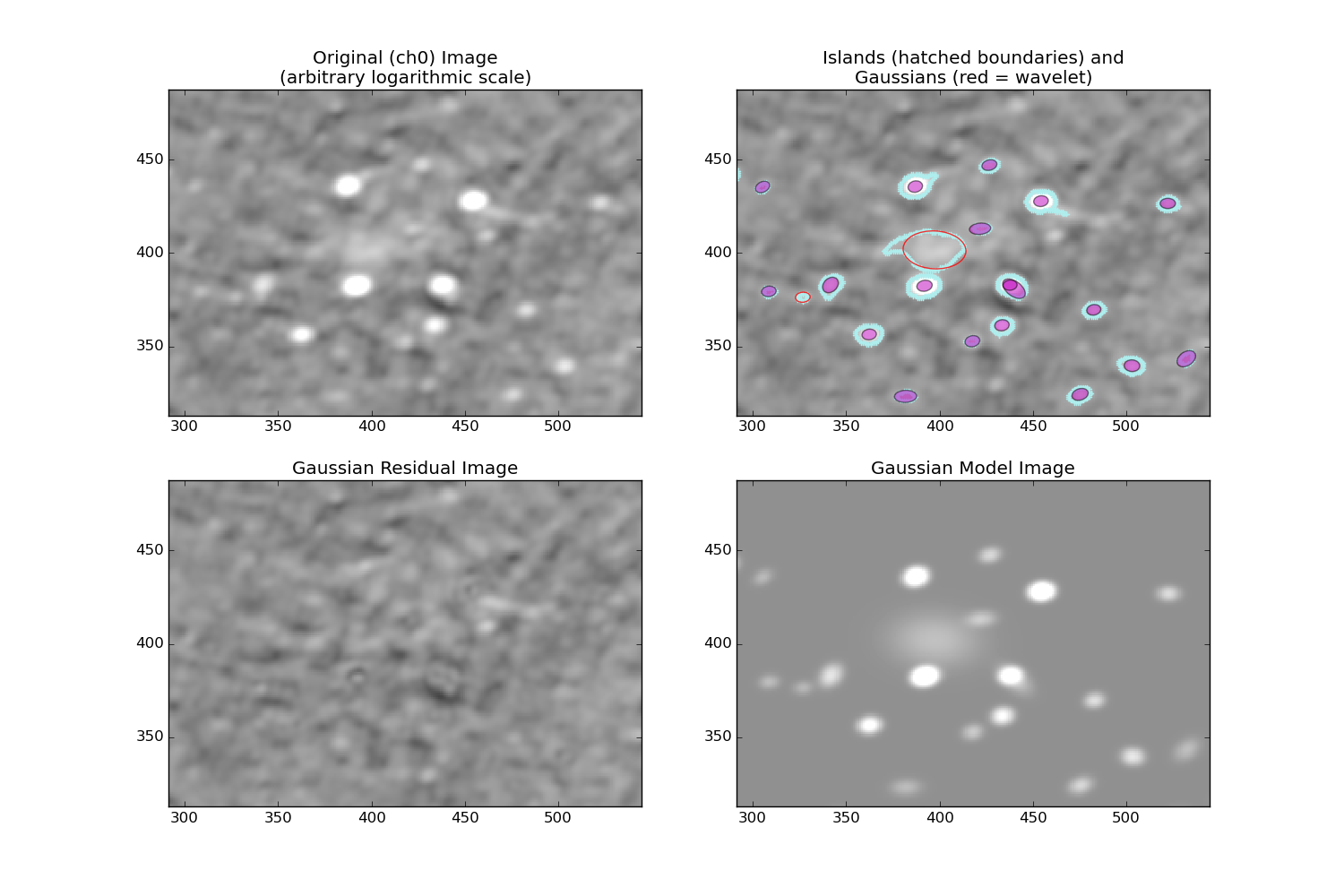}
      \caption{Output of the PyBDSM source finder tool. The axis units are in pixels. From {\it left} to {\it right} and from {\it top} to {\it bottom}: Original image on which PyBDSM is run (in this example, the ID2 map of Table\,\ref{table:fluxes}); islands of significant emission (cyan contours) and fitted Gaussians that model identified sources overlaid on the ID2 map; residual after the subtraction of the fitted sources; model of the reconstructed sky. The big red ellipse corresponds to the diffuse emission in the centre of the cluster.
             }
         \label{fig:pybdsm}
\end{figure*}

The use of packages for the automatic detection of radio sources is becoming more and more necessary and common, especially in the era of the new generation of deep and wide radio surveys that will ultimately bring to the SKA a revolutionary view of the long-wavelength sky \citep[see e.g.][]{2013PASA...30...20N}. In this framework, and in order to get a complementary measure of the total flux density of the diffuse radio source, we ran the automatic source finder PyBDSM \citep{2015ascl.soft02007M} on the tapered images with and without point source subtraction from the uv-data. It is worth noting that PyBDSM finds significant emission in both maps, corresponding to the diffuse radio source that we found by visual inspection and analysis on the maps (see e.g. Fig.\,\ref{fig:pybdsm}). 

We used the threshold technique of the PyBDSM \texttt{process\_image} task\footnote{See \url{http://www.astron.nl/citt/pybdsm/process\_image.html\#general-reduction-parameters} for detailed instructions of \texttt{process\_image}.}, which locates islands of emission above some multiple of the noise in the image (\texttt{thresh\_isl} parameter, set to 3$\sigma$ here). This determines the region where fitting is done. In addition, we set to 5$\sigma$ the source detection threshold in number of sigma above the mean (\texttt{thresh\_pix} parameter). Finally, we activated the wavelet module of \texttt{process\_image} (i.e. \texttt{atrous\_do=True}), which  improves the detection of diffuse sources by doing wavelet transforms at increasing scales of the residual image after subtraction of the initial fitted Gaussians. The modelled Gaussians are shown in the bottom right panel of Fig.\,\ref{fig:pybdsm}, while the top right panel shows the islands of significant emission (cyan) and the position of the fitted Gaussians. Violet ellipses indicate the sources identified on the input radio map (top left), while the red empty ellipse shows the extended source recovered through the wavelet analysis. As shown in the residual maps (bottom left), we nicely managed to fit most of the source components, in particular the central diffuse source. 

The final output catalogs of PyBDSM give  a list of all the Gaussian functions fitted to model the significant emission within the input radio maps and give a source list where different Gaussians are grouped together  if they satisfy objective criteria to be considered as a single source. For each detected source, PyBDSM provides the values of the FWHM of the major and minor axis and the total integrated Stokes I flux density. To subtract the contribution of point sources in the ID2ss case reported in Table\,\ref{table:fluxes}, we ran PyBDSM on the full-resolution Block 3 image of the cluster. The flux densities of compact sources obtained in this way and contained within the cyan region of Fig.\,\ref{fig:pybdsm} were then subtracted from the total flux density of the diffuse source obtained in the ID 2 case. 

The size and flux density of the newly detected radio halo are given as an output of PyBDSM, being 5.76 arcmin $\times$ 4.34 arcmin (914 kpc $\times$ 689 kpc) and 4.91 $\pm$0.04 mJy for the map with ID1. In the case of ID2 and ID2ss, we obtained a halo size of 4.48 arcmin $\times$ 2.70 arcmin, and flux densities of 4.08 $\pm$ 0.07 mJy and 3.72 $\pm$ 0.16 mJy, respectively. 

We note here that, compared to classical measurements performed ``by hand'' (Sect. \ref{sect:char} and Table \ref{table:fluxes}), PyBDSM gives systematically higher values of both source sizes and total flux densities. This is because with classical methods we integrate the surface brightness
of the diffuse source within a region delimited by the 3$\sigma$ contours, while this same region is used by PyBDSM as a support to fit one (or multiple) Gaussian function(s) giving the total flux density of the source(s) whose size is reconstructed based on moment analysis.
 This last method tends to include flux density coming from regions outside the original islands of significant emission, in particular when the wavelet module, which decomposes the residual image that results from the normal fitting of Gaussians into wavelet images of various scales, is activated\footnote{\texttt{atrous\_do} parameter set to \texttt{True}. See \url{http://www.astron.nl/citt/pybdsm/process\_image.html\#a-trous-wavelet-decomposition-module}}.

Both methods are based on different assumptions. In the case of classical ``by-hand'' measurements, for instance, it is assumed that the mean value of the surface brightness of all pixels within 3$\sigma$ contours multiplied by the number of synthesised beams within the considered region is a proper measurements of the integrated flux density for the whole RH. Instead, in the case of PyBDSM, it is assumed that sources are correctly modelled by one or more Gaussian functions of increasing size. We do not conclude that one method is better than the other since they are based on different approaches, but we definitely recommend  not  directly comparing results obtained through a mix of the two different methods  when producing plots such as our Fig.\,\ref{fig:Fighalos}. 

\end{appendix}

\end{document}